\documentclass{aiaa-tc}

 \usepackage{varioref}
 \usepackage{wrapfig}
 \usepackage{threeparttable}
 \usepackage{dcolumn}
  \newcolumntype{d}{D{.}{.}{-1}}
 \usepackage{nomencl}
 \usepackage{subfigure}
 \usepackage{subfigmat}
 \usepackage{fancyvrb}
  \fvset{fontsize=\footnotesize,xleftmargin=2em}
 \usepackage{lettrine}
 \usepackage{hyperref}
 \usepackage{setspace}
 \makenomenclature
 \title{Particle Simulation of Plasma Drag Force Generation in the Magnetic Plasma Deorbit}

 \author{
  Rei Kawashima
  and Junhwi Bak\\
  {\normalsize\itshape The University of Tokyo, Tokyo 113-8656, Japan}\\
  and\\
  Shinji Matsuzawa
  and Takaya Inamori\\
  {\normalsize\itshape Nagoya University, Aichi 464-8603, Japan}
 }

 \AIAApapernumber{200?-????}
 \AIAAconference{Conference Name, Date, and Location}
 \AIAAcopyright{\AIAAcopyrightD{200?}}

\begin{document}

\maketitle

\begin{abstract}
	The magnetic plasma deorbit method has been proposed for micro and nanosatellites in the low earth orbit (LEO).
	The magnetic plasma deorbit utilizes the plasma drag force generated by the interaction between space plasma and the magnetic field surrounding magnetic torquers (MTQs).
	In this study, the plasma drag force generated by the magnetic plasma deorbit method is estimated by using a plasma flow simulation based on the fully kinetic model.
	The dependences of the plasma drag force on the MTQ angle, atmospheric plasma density, and MTQ magnetic moment are investigated.
	The simulation results show that the structures of the bow shock and magnetosphere are formed in the vicinity of the MTQs.
	The predicted plasma drag force is 105.2 nN for an MTQ of 10 A m$^2$ when the MTQ magnetic moment is parallel to the satellite velocity direction.
	The plasma drag force shows a strong dependence on the magnetic moment angle.
	In addition, the plasma drag force is proportional to the atmospheric plasma density and the MTQ magnetic moment.
	Steady drag force generation in the magnetic plasma deorbit is validated, and the sensitivities of the drag force to the relevant parameters are elucidated.
\end{abstract}

\printnomenclature

\nomenclature[1m]{$m$}{mass}
\nomenclature[1u]{$u$}{velocity}
\nomenclature[1v]{$v$}{particle speed}
\nomenclature[1f]{$F$}{drag force}
\nomenclature[1cd]{$C_{\rm d}$}{drag coefficient}
\nomenclature[1l]{$L$}{magnetosphere size}
\nomenclature[1A]{$A$}{magnetic vector potential}
\nomenclature[1s]{$S$}{face area}
\nomenclature[1a]{$a$}{semi-major axis}
\nomenclature[1b]{$B$}{magnetic field}
\nomenclature[1e]{$E$}{electric field}
\nomenclature[1e]{$e$}{elementary charge}
\nomenclature[1q]{$q$}{particle charge}
\nomenclature[1kb]{$k_{\rm B}$}{Boltzmann constant}
\nomenclature[1rl]{$r_{\rm L}$}{Larmor radius}
\nomenclature[1md]{$M_{\rm d}$}{magnetic moment}
\nomenclature[1h]{$H$}{fitting parameter}
\nomenclature[1n]{$n$}{number density}
\nomenclature[1td]{$t_{\rm deorbit}$}{deorbit duration}
\nomenclature[1h]{$h$}{altitude}
\nomenclature[1fm]{$f_{\rm m}$}{artificial electron mass coefficient}
\nomenclature[1j]{$j$}{current density}
\nomenclature[1u]{$u$}{velocity}
\nomenclature[2f]{$\phi$}{space potential}
\nomenclature[2r]{$\rho$}{mass density}
\nomenclature[2c]{$\Gamma$}{number flux density}
\nomenclature[2c]{$\gamma$}{standard gravitational parameter}
\nomenclature[2t]{$\theta$}{angle of the magnetic moment}
\nomenclature[2e]{$\varepsilon$}{vacuum permittivity}
\nomenclature[2m]{$\mu$}{vacuum permeability}
\nomenclature[be]{e}{electron}
\nomenclature[bemp]{emp}{empirical}
\nomenclature[bp]{p}{plasma}
\nomenclature[bi]{i}{ion}
\nomenclature[bs]{s}{species}
\nomenclature[bc]{c}{critical}
\nomenclature[bm]{MTQ}{magnetic torquer}
\nomenclature[bae]{aero}{aerodynamic}
\nomenclature[bind]{ind}{induced magnetic field}
\nomenclature[bini]{ini}{initial altitude}
\nomenclature[bref]{ref}{reference}
\nomenclature[bsim]{sim}{simulated}
\nomenclature[bE]{E}{Earth}
\nomenclature[bs]{sat}{satellite}
\nomenclature[ba]{0}{uniform flow}
\nomenclature[bz]{$\parallel$}{magnetic moment parallel to the satellite velocity}
\nomenclature[bz]{$\perp$}{magnetic moment perpendicular to the satellite velocity}

\section{Introduction}
	Because the development of nano and microsatellites is fast and low-cost, a number of them were successfully launched for various types of missions.
	Most of these satellites are used in the low earth orbit (LEO).
	In general, these satellites complete their missions within ten years; furthermore it must be noted that many of them do not stay in orbit for very long periods of time.
	The lifetime of CubeSat-class satellites is less than 200 days on average [\citen{SwartwoutJoSS2013}].
	There also exist long-lifetime nanosatellites.
	An example of a long-lifetime nanosatellite is CubeSat XI-V [\citen{Funase:2007aa}].
	This CubeSat was launched in October 2005, and it is still operational in December 2017.
	One of the reasons for the reduced lifetime of nanosatellites is inadequate thermal or electronic designs.
	If the technologies of nanosatellite development advance further, the average lifetime of nanosatellites is expected to increase in the future.
	However, the increasing development and use of nanosatellites will inevitably cause the issue of space debris.
	Propulsion systems are generally not installed on nanosatellites because of strict constraints on mass and volume.
	Hence, deorbit methods for nano and microsatellites have been of concern in recent years to prevent the accumulation of space debris.

\begin{figure}[t]
	   \begin{center}
   		\includegraphics[width=80mm]{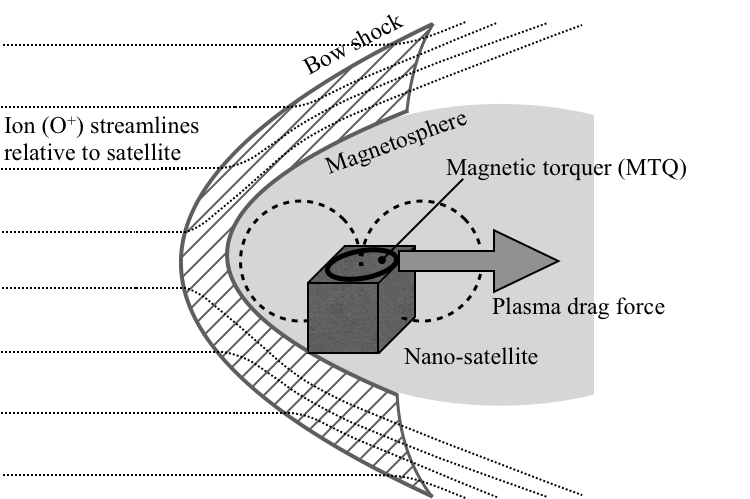}
		\end{center}
   	\caption{Concept of the drag force generation in the magnetic plasma deorbit method.
		}
   	\label{fig:concept}
   \end{figure}

	As a deorbit method for nano and microsatellites, a method referred to as the magnetic plasma deorbit has been proposed by the authors [\citen{Inamori2015192}].
	The concept of the method is illustrated in Fig. \ref{fig:concept}.
	This method uses the plasma drag force generated by the interaction between space plasma and the magnetic field surrounding a satellite.
	The magnetic field is generated by a magnetic torquer (MTQ), which is generally installed on nanosatellites for attitude control.
	Thus, the advantage of this method is that it requires no additional equipment such as propulsion systems or large structures for deorbit.

	The fundamental physics of the drag force generation in the magnetic plasma deorbit is supposed to be similar to that of magnetic sails [\citen{ZUBRIN:1991aa}].
	In the research of magnetic sails, Funaki et al. proposed an analytical model for the prediction of the plasma drag force as follows [\citen{Funaki2007}]:
	\begin{equation}
		F=C_{\rm d,p}\frac{1}{2}m_{\rm i}n_0u_{\rm 0}^2\pi L^2,
		\label{eq:model1}
	\end{equation}
	where $L$ is the characteristic length of the magnetosphere, and it is estimated by considering the balance of magnetic and plasma pressures as follows:
	\begin{equation}
		L=\left(\frac{M_{\rm d}^2}{8\mu\pi^2n_0m_{\rm i}u_{\rm sat}^2}\right)^{1/6}.
		\label{eq:model2}
	\end{equation}
	In Ref. [\citen{Inamori2015192}], to evaluate the plasma drag force generated in the magnetic plasma deorbit method, Eqs. (\ref{eq:model1}) and (\ref{eq:model2}) are used.
	However, this analytic model was used for magnetic sails where solar wind plasma was involved, and the model was not validated for the magnetic plasma deorbit in LEO plasma.
	It is noted that the concept of the magnetic plasma deorbit in Fig. \ref{fig:concept} is made based on the plasma flow characteristics of magnetic sails.
	Thus, it is also unknown whether various structures, such as the bow shock and magnetosphere, are formed in the case of the magnetic plasma deorbit.

	Similarly, the sensitivities of the plasma drag force in the magnetic plasma deorbit to the relevant parameters were not elucidated.
	First, the effect of the magnetic moment angle $\theta$ on the plasma drag force is concerned.
	In the case of magnetic sails, it is known that the $\theta$ has an influence on the plasma drag force [\citen{Nishida:2012aa}], and hence $\theta$ is supposed to affect the performance of the magnetic plasma deorbit.
	Moreover, the effect of the atmospheric plasma density $n_{0}$ and magnetic moment of an MTQ $M_{\rm d}$ is concerned, since these parameters vary in wide ranges in the cases of nanosatellites in the LEO.
	The aforementioned model yields the relations of $F\propto n_{\rm 0}^{2/3}$ and $F\propto M_{\rm d}^{2/3}$.
	The validity of these relations must be checked in the case of the magnetic plasma deorbit.
	
	In this study, a plasma flow simulation based on the fully kinetic model is conducted to analyze the plasma drag force generated in the magnetic plasma deorbit.
	The primary objective is to confirm steady plasma drag force generation in the magnetic plasma deorbit method.
	Further, the simulated plasma drag force is compared with that of the analytic model in Eqs. (\ref{eq:model1}) and (\ref{eq:model2}), and the validity of the model is examined.
	Additionally, a parametric study is performed to investigate the effect of $\theta$, $n_0$, and $M_{\rm d}$ on the plasma drag force.
	Concerning the relationships between the plasma drag force and these parameters, the simulation results are compared with those of the analytic model.

\section{Basics of the Electrostatic Fully Kinetic Model}

\subsection{Plasma Flow}
	The fully kinetic particle model is used for the description of the magnetized plasma flow.
	The motions of ions and electrons are computed by using the two-dimensional (2D) three-velocity particle-in-cell (PIC) method.
	Several millions of particles are bundled into one macroparticle, and each macroparticle is accelerated by outer forces.
	The equation of motion for each macroparticle is formulated as follows:
   \begin{equation}
      m_{\rm s}\frac{d \vec{v}_{\rm s}}{d t}=\vec{F}_{\rm s},
      \label{eq:particle1}
   \end{equation}
   \begin{equation}
      \frac{d \vec{x}_{\rm s}}{d t}=\vec{v}_{\rm s}.
      \label{eq:particle2}
   \end{equation}
   $\vec{F}_{\rm s}$ is the electromagnetic (EM) force and subscript s denotes each species.
	It is noted that $\vec{v}$ is the relative speed to the satellite as depicted in Fig. \ref{fig:concept}.
   The EM force acting on each macroparticle can be expressed as follows:
   \begin{equation}
      \vec{F}_{\rm s}=q\left(\vec{E}+\vec{v}_{\rm s}\times\vec{B}\right),
      \label{eq:force}
   \end{equation}
   where $q$ is equal to $e$ for ions and $-e$ for electrons. 
   The magnetic field is calculated by the superposition of the MTQ magnetic field and induced magnetic field: $\vec{B}=\vec{B}_{\rm MTQ}+\vec{B}_{\rm ind}$.
   
   An appropriate plasma flow model was discussed in previous researches on magnetic sails [\citen{Kajimura:2012aa,Ashida:2013aa}].
   The criterion for determining the plasma flow model is the ratio of the Larmor radius to the magnetosphere size $L$.
   In the case of LEO plasma and an MTQ strength of 1--10 A m$^2$, the ratio of the ion Larmor radius to $L$ is supposed to be on the order of 10$^3$, and the ratio of the electron Larmor radius to $L$ is on the order of 1.
   In this case, the kinetic model should be used for both ions and electrons.
   Thus, the fully kinetic model is used in the modeling of the LEO plasma flow surrounding the nanosatellite MTQs.   
   Interparticle collisions are neglected in the present model.
   All of the mean-free paths of ion--ion Coulomb, electron--ion Coulomb, ion--neutral momentum-exchange, and electron--neutral momentum-exchange collisions are estimated to be greater than 10$^3$ m under atmospheric conditions at high altitudes where the magnetic plasma deorbit method is used.
   Very few particles experience collisions at the spatial scale of the magnetic plasma deorbit since it is 1 m as shown later.

\subsection{EM Field}
	In this study, the electric field is assumed electrostatic, and the space potential is obtained through Gauss's law.
   In the 2D simulation in the $x-y$ plane, the uniformity of space potential is assumed in the $z$-direction.
	The 2D Poisson equation of Gauss's law is formulated as follows:
	\begin{equation}
      \nabla\cdot\left(-\varepsilon\nabla\phi\right)=e\left(n_{\rm i}-n_{\rm e}\right),
      \label{eq:poisson}
   \end{equation}
   After deriving the space potential, the electric field is calculated by the relation: $\vec{E}=-\nabla\phi$.
   
	The magnetic field is also assumed static, and it is calculated through the vector potentials of the MTQ and induced magnetic field.
	Concerning the MTQ, a line dipole is assumed for the 2D simulation in the $x-y$ plane.
	The vector potential of a line dipole is expressed as follows:
	\begin{equation}
		A_{{\rm MTQ},z}=\frac{\mu}{2\pi}M_{\rm d}\left(-x\sin\theta+y\cos\theta\right)\left(x^2+y^2\right)^{-1},
		\label{eq:dipole}
	\end{equation}
	where the definition of the $x-y$ coordinate and $\theta$ are as illustrated in Fig. \ref{fig:condition}.
   The vector potential for the induced magnetic field is calculated by using Poisson's equation for a vector potential as follows:
   \begin{equation}
      \nabla\left(-\nabla\cdot A_{{\rm ind},z}\right)=\mu\left(j_{{\rm i},z}-j_{{\rm e},z}\right),
      \label{eq:vecpot}
   \end{equation}
   Note that the current in the $z$-direction is induced by the effects of magnetization. 
   As discussed in Sec. IV.B., ions are not magnetized with the magnetic field strength concerned in the magnetic plasma deorbit method, and hence the induced magnetic field is mainly caused by the electron current.
   After deriving the magnetic vector potentials, the magnetic field in the $x-y$ plane is calculated as follows:   
   \begin{eqnarray}
      B_x=\ \ \frac{\partial}{\partial y}\left(A_{{\rm MTQ},z}+A_{{\rm ind},z}\right),\\
      B_y=-\frac{\partial}{\partial x}\left(A_{{\rm MTQ},z}+A_{{\rm ind},z}\right).
      \label{eq:induced}
   \end{eqnarray}
   
   In the present model, the assumed electrostatic and magnetostatic fields that do not include the effects of displacement current.
   Here, we call this model the electrostatic--magnetostatic (ESMS) model.
   Information about the displacement current propagates in space with the speed of light, which means almost instantaneously at the time scale of the ion flow.
   Note that the ratio of the speed of light to the speed of the ion flow surrounding a satellite in the LEO (i.e., condition number) is $\sim 4\times10^{4}$.
   The time derivative of the electric field would be small at the ion time scale, and the ESMS model is supposed to be a good approximation for the simulation of the magnetic plasma deorbit method.
   Another approach to obtaining the electric and magnetic fields is to directly solve Faraday's law and Ampere's law in Maxwell's equations [\citen{Harned:1982aa}].
   The model is called the EM model, and it was used in the full particle and particle-fluid hybrid modeling of magnetic sails [\citen{Ashida:2013aa,Kajimura:2010aa}].
   It was reported that the ESMS model yields almost the same thrust as the EM model in the full particle simulations of a 2D magnetic sail, for which the condition number is 600 [\citen{AshidaThesis}].
   This fact supports the validity of the ESMS model for steady-state simulations of the magnetic plasma deorbit method.
   
   Further, the Earth geomagnetic field (GMF) is neglected in this study.
	In the research of magnetic sails, Nishida and Funaki conducted a numerical simulation of a magnetic sail with the interplanetary magnetic field (IMF) and observed that the IMF affects the thrust characteristics [\citen{Nishida:2012aa}].
	It was reported that the IMF enhances the thrust and induces a pitching moment on the magnetic sail.
	The detailed effect of the background magnetic field is complicated even in the sense of qualitative trends since two angles of the satellite magnetic moment and IMF are involved.
	The GMF is also supposed to influence the performances of the magnetic plasma deorbit method.
	In the case of the magnetic plasma deorbit method used in the LEO, the direction of the GMF changes depending on the position of satellites, i.e., longitude and latitude.
	Hence, a detailed analysis with various combinations of the MTQ and GMF fields is required to understand the GMF effect comprehensively.
	In this paper, we focus on the case of no GMF for simplicity to benchmark the basic characteristics of the magnetic plasma deorbit method.

\subsection{Artificial Electron Mass Model}
	\label{sec:aem}
	In particle simulations of plasma, there are mainly three restrictions for the time step: 1) resolution of the plasma frequency, 2) Courant--Friedrichs--Lewy (CFL) condition for electron velocity, and 3) CFL condition for ion velocity.
	Here, the CFL conditions are related to the speed of information propagation in electrostatic electron or ion waves [\citen{chen1984introduction}].
	Among the restrictions above, the CFL condition for electron velocity typically requires the smallest time step in the case of the LEO plasma flow.
	The use of a small time step results in a large number of time steps to achieve the steady state at the time scale of ions.
	This issue of numerical stiffness arising from a large mass ratio of $m_{\rm i}/m_{\rm e}$ appears in many full PIC simulations, and concessions must be made to speed up the convergence [\citen{ChoPoP2013,Szabo:2013aa}].
	In the present modeling, to accelerate the simulation, the artificial electron mass (AEM) model is used.
	In the AEM model, the electron mass is artificially increased as follows:
   \begin{equation}
   	m_{\rm e}\rightarrow m_{\rm e}'=f_{\rm m}m_{\rm e}.
		\label{eq:amr1}
   \end{equation}
	Increasing the electron mass by a factor $f_{\rm m}$ reduces the electron velocity by $\sqrt{f_{\rm m}}$ and accelerates the convergence by $\sqrt{f_{\rm m}}$.
	
	The magnetic plasma deorbit uses charge separation caused by the difference in the Larmor radii between ions and electrons.
	Thus, the electron gyroradius must be maintained while $f_{\rm m}$ is introduced.
	The electron gyroradius is expressed as follows:
	\begin{equation}
		r_{\rm L,e}=\frac{m_{\rm e}v_{\rm e}}{eB_{\rm e}}.
	\end{equation}
	In this simulation, the magnetic flux density for electrons is increased as follows:
   \begin{equation}
   	B_{\rm e}\rightarrow B_{\rm e}'=f_{\rm m}B_{\rm e}.
		\label{eq:amr2}
   \end{equation}
   The AEM factor is selected as $f_{\rm m}=200$ for all the simulation cases in this study.
   The validity of the selection of this value is discussed in the appendix.
   After the application of the AEM model, the time step is typically set to $\Delta t= 10^{-8}$ s.

\section{Calculation Condition and Numerical Method}
\subsection{Calculation Condition}
\begin{figure}[t]
   	\begin{center}
	   	\includegraphics[width=65 mm]{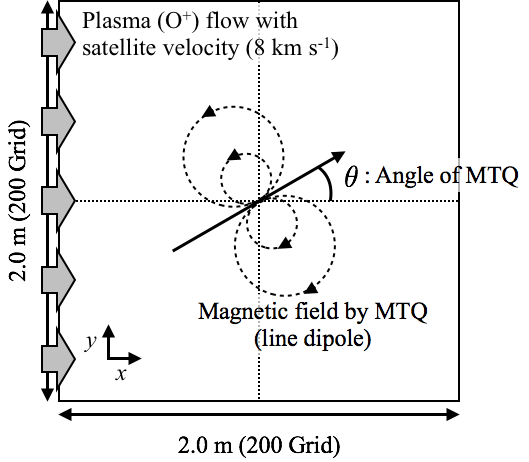}
   	\end{center}
		\vspace{-2mm}
   	\caption{Calculation condition of the magnetic plasma deorbit.
			The control volume surrounding a satellite is used.}
   	\label{fig:condition}
   \end{figure}

	In the present study, satellites at an altitude of 300--1000 km are concerned.
	The plasma environment in this altitude range is referred to as the exosphere [\citen{SchunkBook2004}].
	This region is inside the magnetopause generated by the geomagnetic field, and the effect of high-velocity solar wind is small.
	Table 1 lists the assumed parameters for the simulation of the magnetic plasma deorbit in the LEO.
	The properties of plasma are calculated through the International Reference Ionosphere (IRI) 2012 model as functions of the altitude and time [\citen{Bilitza}].
	In this simulation, the altitude of $h=800$ km and daytime are assumed as the calculation conditions.
	All ions are assumed to be O$^+$.
	This is because the fraction of O$^+$ is greater than 0.9 for both daytime and nighttime, and the contributions of other ion species to the drag force are supposed to be small.
	The ion and electron temperatures are also obtained from the IRI model.
	The satellite velocity is 8,000 m s$^{-1}$.
	In the simulation, the ion flow velocity is assumed to be equal to the satellite velocity in the control volume surrounding a satellite.
	In the exosphere region, the plasma wind tends to blow horizontally from the subsolar heated side to the night cold side.
	The typical wind speed ranges from 100 m/s to 300 m/s for quiet geomagnetic conditions [\citen{SchunkBook2004}].
	Since the wind speed is smaller than the satellite velocity by one order of magnitude, the effect of plasma wind is neglected in the present study.
	The calculation domain is illustrated in Fig. \ref{fig:condition}.
	The control volume surrounding the satellite is used, and the plasma flow is introduced to the domain from the left-hand side boundary.
	It is known that the typical satellite velocity is larger than the ion thermal velocity in the LEO environments [\citen{HastingsJGR1995}].
	Hence, the number flux density of ions from the left-hand side boundary is given by using the satellite velocity as $\Gamma_{\rm i}=n_{\rm 0}u_{\rm sat}$.
	However, for the top and bottom boundaries, the number flux density of ions is given by a random flux.
	The random flux can be written as follows:
	\begin{equation}
		\Gamma_{\rm s}=\frac{1}{4}n_{\rm s}\sqrt{\frac{8k_{\rm B}T_{\rm s}}{\pi m_{\rm s}}},
	\end{equation}
	where ${\rm s}$ denotes each species.
	Concerning electrons, the thermal velocity is much greater than the satellite velocity.
	Thus, the number flux density of electrons based on the random flux is introduced to the calculation domain from the left, right, top, and bottom boundaries.
	Because the satellite velocity is much smaller than the speed of light, the Lorentz transformation is not considered in this calculation.
	A satellite is located at the center of the calculation field, and the dipole magnetic field is generated around the satellite.

	\begin{table}[t]
		\begin{center}
			\caption{Parameters assumed for the simulation of the magnetic plasma deorbit in the LEO.}
			\begin{tabular}{lcc}
   			\Hline
				 Parameter & Symbol & Value\\ \hline
			    Ion number density, m$^{-3}$ & $n_{\rm 0}$ & 1.0 $\times$ 10$^{11}$\\
				 Ion species & -- & O$^+$\\
				 Ion temperature, K & $T_{\rm i}$ & 1230\\
				 Electron temperature, K & $T_{\rm e}$ & 2000\\
				 Satellite velocity, m s$^{-1}$ & $u_{\rm sat}$ & 8000\\
				 Magnetic moment, A m$^2$  & $M_{\rm d}$ & 10\\
				 Angle of MTQ & $\theta$ & 0--$\pi/2$\\
				 \Hline
			\end{tabular}
      	\label{tab:parameter}
		\end{center}
	\end{table}

\subsection{Numerical Method}
	\label{sec:method}
	
	The MTQ magnetic field is input at the beginning of the simulation.
	In addition, one million macroparticles of ions and electrons are uniformly distributed in the calculation region.
	The basic computational flow in the time marching consists of the electrostatic and magnetic field calculations and particle motion calculations.
	Possion's equation in Eq. (\ref{eq:poisson}) is solved for the scalar potential by applying the second-order central differencing.
	An iterative method of successive line over-relaxation method is used with a truncation error of 10$^{-8}$.
	After deriving the scalar potential, the electric field is calculated by applying the second-order central differencing.
	The vector potential and the magnetic field are also calculated by applying the second-order central differencing to Eqs. (\ref{eq:vecpot}) and (\ref{eq:induced}).
	
	The overall time advancement is implemented in an explicit manner.
	The time advancement of particle motions described by Eqs. (\ref{eq:particle1}) and (\ref{eq:particle2}) is computed by using the Buneman--Boris method [\citen{BirdsallBook2004}].
	After the motion calculations of ions and electrons, the charge density and the current density are calculated by weighting the particle mass and momentum to each cell.
	The weighting function in the PIC method is the second-order spline function.
	The number of macroparticles for each species in the calculation field is normally one million, which corresponds to 25 particles per cell.
	The inflow of particles from the boundaries is implemented by assuming the Maxwellian particle energy distribution with the ion and electron temperatures listed in Table \ref{tab:parameter}.
	The plasma drag force is computed by integrating the momentum changes of particles in the x-direction, while they traverse the calculation field.

	In non-neutral plasma simulations, the grid size $\Delta x$ should be determined to satisfy the relation $\Delta x/\lambda_{\rm D}\simeq 1$,
	where $\lambda_{\rm D}$ is the Debye length [\citen{MASON1981233}].
	In this simulation, the grid sizing of $\Delta x\simeq 1.0\times10^{-2}$ m is used, which corresponds to $\Delta x/\lambda_{\rm D}\simeq 1$ when the parameters listed in Table \ref{tab:parameter} are used.

\section{Results and Discussion}
\subsection{Confirmation of Steady Drag Force Generation}
	\label{sec:force}
\begin{figure}[b]
		\begin{center}
   		\includegraphics[width=70mm]{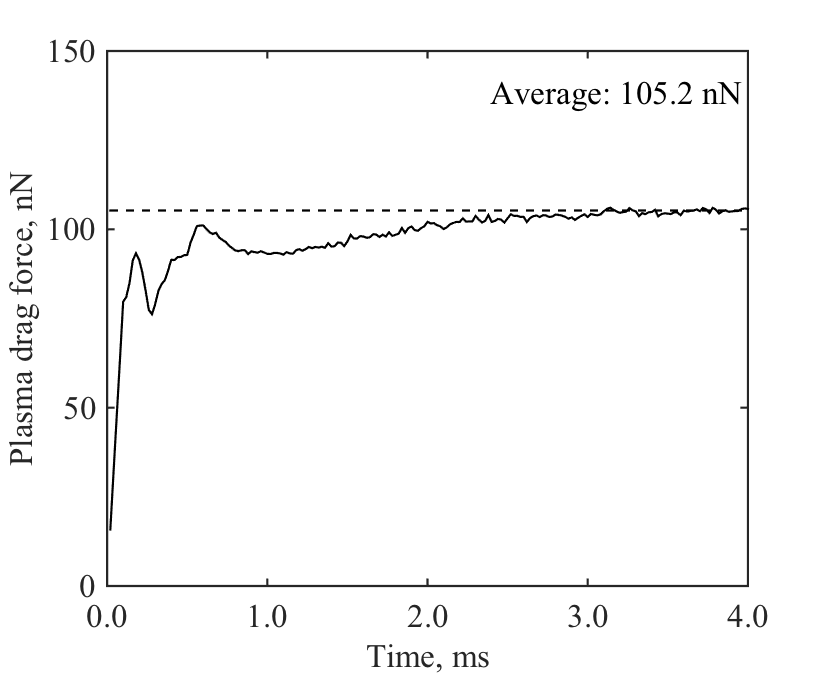}
		\end{center}
		\vspace{-2mm}
   	\caption{Time history of the plasma drag force.}
   	\label{fig:force_time}
   \end{figure}

	After several millions of time steps, the steady state is confirmed in the plasma flow.
	A typical iteration history of the calculated plasma drag force is shown in Fig. \ref{fig:force_time}.
	Here, the case of $n_0$ = 10$^{11}$ m$^{-3}$, $M_{\rm d}$ = 10 A m$^2$, and $\theta$ = 0 is simulated.
	In this case, the quasi-steady state is reached at $t\geq 3.0$ ms with the tolerable statistical noise stemming from the particle model.
	The steady plasma drag force is estimated by averaging the history for the last 0.5 ms.
	In what follows, this time-averaged steady drag force is discussed in each case.

   \begin{figure}[t]
		\begin{center}
   		\includegraphics[width=80mm]{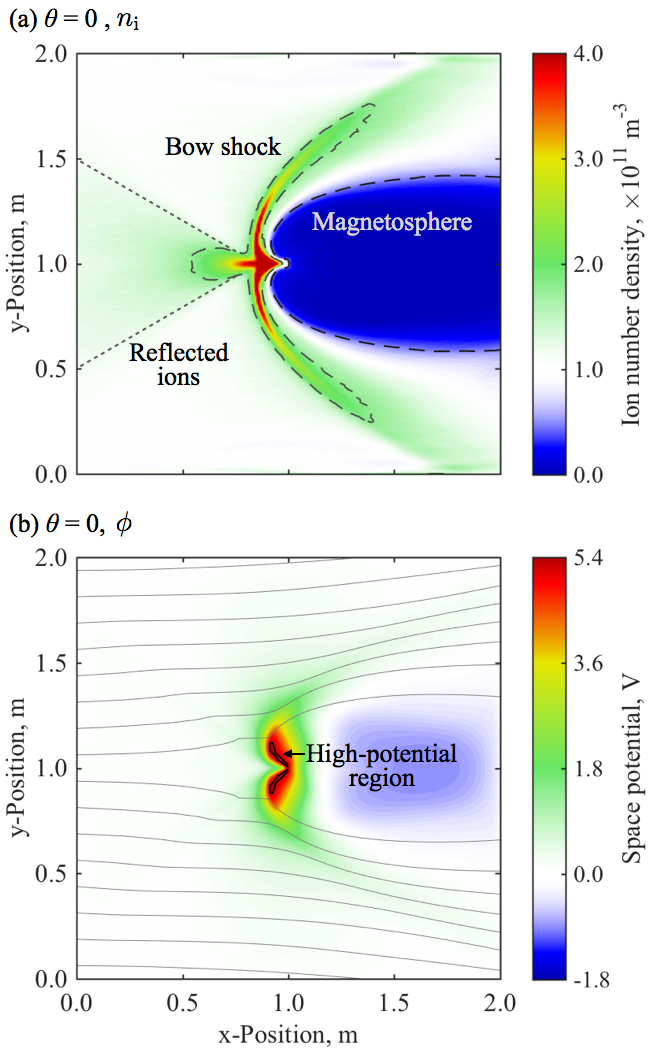}
		\end{center}
		\vspace{-2mm}
   	\caption{2D distributions of the ion number density and space potential in the case of $\theta$ = 0.}
   	\label{fig:dist2d_1}
   \end{figure}
   \begin{figure}[b]
   	\begin{center}
	   	\includegraphics[width=70mm]{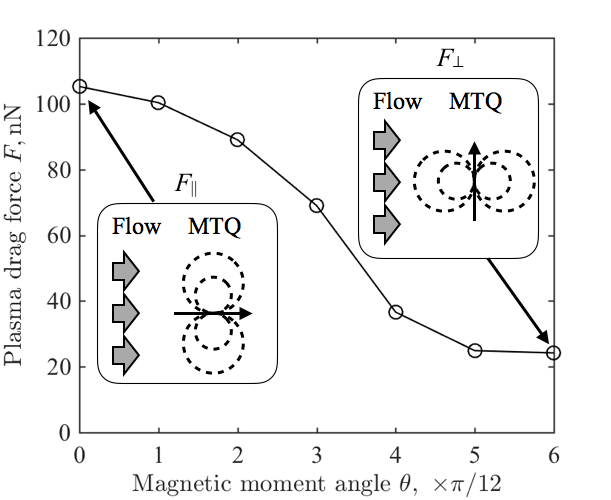}
   	\end{center}
		\vspace{-2mm}
   	\caption{Dependence of the plasma drag force on the magnetic moment angle.}
   	\label{fig:ang}
   \end{figure}

	The 2D distributions of the ion number density and space potential are shown in Fig. \ref{fig:dist2d_1}.
	The bow shock and magnetosphere are observed as shown in Fig. \ref{fig:dist2d_1}(a).
	Here, the boundaries of the bow shock and magnetosphere are defined by the contour lines of $n_{\rm i}=2.0n_{\rm 0}$ and $n_{\rm i}=0.5n_{\rm 0}$, respectively.
	The coefficient of 0.5 is chosen for clear visualization of the boundary.
	As shown in Sec. IV.B., the ion number density steeply changes at the boundary of the magnetosphere.
	Thus, even if this coefficient is changed between 0.2 and 0.8, the location of the boundary is almost unchanged.
	As shown in Fig. \ref{fig:dist2d_1}(b), the high-potential region is formed at the upstream of the MTQ center.
	The high-potential region is formed by the charge separation between the ion and electron flows, and this charge separation is generated owing to the magnetization of electrons by the MTQ magnetic field.
	The high-potential region brakes and deflects ions approaching the MTQ center.
	The structures of the bow shock and magnetosphere are formed as consequences of the ion braking.
	As a conclusion, the drag force generation mechanism as shown in Fig. 1 is confirmed by the simulation.
	
	In the case of the result in Fig. \ref{fig:force_time}, the simulated steady plasma drag force is 105.2 nN.
	If one substitutes this plasma drag force into the analytical model in Eq. (\ref{eq:model1}), the corresponding drag coefficient is calculated as $C_{\rm d,p}= 0.2$.
	In the experiments and simulations of a magnetic sail, the drag coefficient is estimated as 0.5--2.0 [\citen{Funaki2007,Nishida:2006aa}].
	The very small $C_{\rm d,p}$ in the magnetic plasma deorbit is attributed to the small magnetosphere size $L$.
	Ashida et al. showed that $C_{\rm d,p}$ is no longer constant against $L$ in the cases of small magnetospheres, in which the ratio of the ion Larmor radius to the magnetosphere size, $r_{\rm L,i}/L$, is greater than unity [\citen{Ashida:2013aa}].
	Under the calculation condition of the magnetic plasma deorbit, the ratio of $r_{\rm L,i}/L$ is approximately 10$^3$.
	Hence, the analytic model of magnetic sails is not valid for the prediction of the plasma drag force in the magnetic plasma deorbit method.	
	The validity of the simulated plasma drag forces is discussed in Appendix B via a numerical simulation of a small-scale magnetic sail.
	
\subsection{Dependence of the Plasma Drag Force on the Magnetic Moment Angle of an MTQ}
	\label{sec:ang}
	Figure \ref{fig:ang} shows the variation of the plasma drag force $F$ when the MTQ magnetic moment angle $\theta$ is changed from 0 to $\pi/2$ with an increment of $\pi/12$.
	The maximum $F$ is observed when $\theta = 0$, i.e., the MTQ magnetic moment is parallel to the direction of the satellite velocity.
	$F$ decreases as $\theta$ increases, and the minimum $F$ is generated when $\theta = \pi/2$, i.e., the MTQ magnetic moment is perpendicular to the direction of the satellite velocity.
	In what follows, the plasma drag forces in the cases of $\theta = 0$ and $\theta = \pi/2$ are referred to as $F_{||}$ and $F_{\perp}$, respectively.
	The similar trend of $F_{||}>F_{\perp}$ was also confirmed in the magnetohydrodynamic analysis of a magnetic sail [\citen{Nishida:2012aa}].
	Therefore, the dependence of $F$ on $\theta$ is qualitatively consistent with that of the magnetic sail.

	However, in the quantitative sense, the dependence of $F$ on $\theta$ does not agree with that of the magnetic sail.
	In the case of the magnetic plasma deorbit, $F_{||}$ and $F_{\perp}$ are 105.2 nN and 24.2 nN, respectively, and the ratio of these drag forces is about 3.5.
	However, this ratio is approximately 1.5 in the case of a magnetic sail [\citen{Nishida:2006aa}].
	The plasma drag force generated in the magnetic plasma deorbit shows a strong dependence on the magnetic moment angle, compared with the magnetic sails.
	One of the possible reasons is the assumption of the dipole magnetic field in this simulation.
	In the magnetic sail simulation in Ref. [\citen{Nishida:2006aa}], the initial magnetic field was generated by a ring coil.
	If the magnetic field is generated by ring coils, finite-coil radius effects appear and may change the dependence of the drag force on the magnetic moment angle.

\begin{figure}[t]
		\begin{center}
   		\includegraphics[width=80mm]{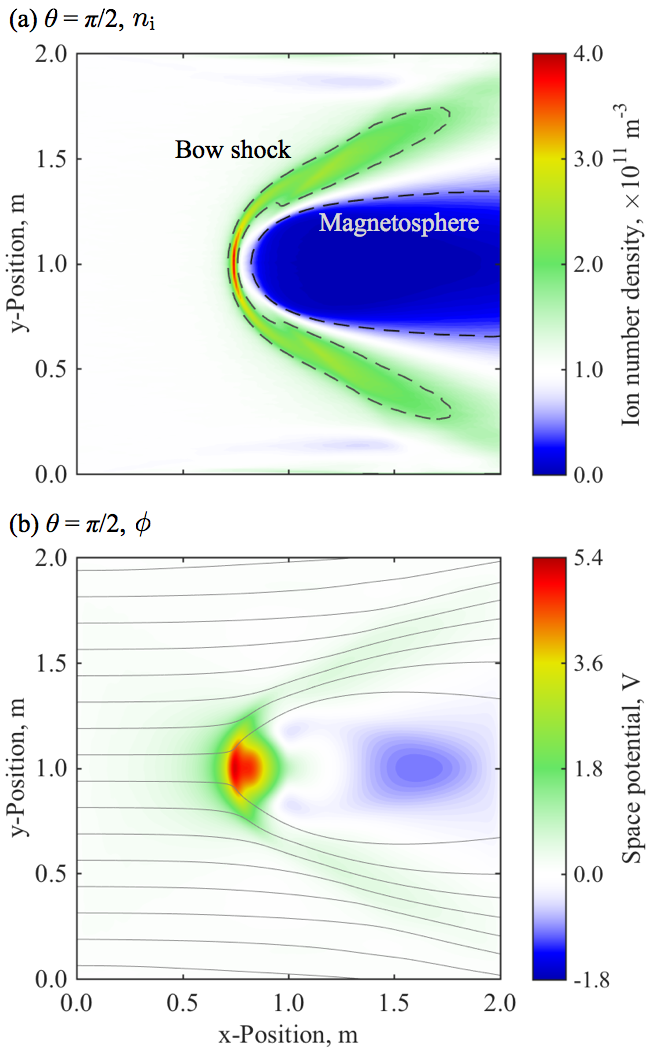}
		\end{center}
		\vspace{-2mm}
   	\caption{2D distributions of the ion number density and space potential in the cases of $\theta$ = $\pi/2$.}
   	\label{fig:dist2d_2}
   \end{figure}

	Another possible cause of the strong dependence on the magnetic moment angle is considered using the 2D distributions.
	The steady-state distributions of the ion number density and space potential in the case of $\theta=\pi/2$ are shown in Fig. \ref{fig:dist2d_2}.
	In the analytical model in Eq. (\ref{eq:model1}), the plasma drag force is associated with the magnetosphere size $L$.
	However, $L$ is obtained by Eq. (\ref{eq:model1}) as $L\sim 1.5$ m in the case of $M_{\rm d}=10$ A m$^2$, which is much larger than the magnetosphere sizes observed in this simulation.
	Hence, another definition is used for the magnetosphere size to reflect the simulation results.
	In Figs. \ref{fig:dist2d_1} and \ref{fig:dist2d_2}, the boundary of the magnetosphere is defined by the contour lines of $n_{\rm i}=0.5\times10^{11}$ m$^{-3}$. 
	Based on the simulation results, $L_{\rm sim}$ is defined as the magnetosphere size in the $y$-direction at $x=1.0$ m.
	$L_{\rm sim}$ is estimated as 0.44 m and 0.40 m in the cases of $\theta$ of 0 and $\pi/2$, respectively. 	
	The difference in $L_{\rm sim}$ is not so large as the difference between $F_{||}$ and $F_{\perp}$.
	Thus, the difference in the magnetosphere size cannot be the reason for the strong dependence on the magnetic moment angle of the magnetic plasma deorbit.
	
	In the case of $\theta=0$, the ion reflection from the center of the MTQ to the $-x$-direction is observed, as shown in Fig. \ref{fig:dist2d_1}(a).
	The reflected ions are not observed in the case of $\theta=\pi/2$.
	The reflected ions may contribute to the strong plasma drag force of $F_{||}$.
	The drag force generated by the reflected ions can be roughly estimated by using the $\phi$ distribution.
	The critical space potential $\phi_{\rm c}$, which is required for the reflection of incident ions, can be calculated as follows:
	\begin{equation}
		\frac{1}{2}m_{\rm i}u_{\rm sat}^2=e\phi_{\rm c}.
		\label{eq:phic}
	\end{equation}
	In the case of $u_{\rm sat}$ = 8.0 km s$^{-1}$, $\phi_{\rm c}$ is calculated as 5.3 V.
   
	The region where $\phi > \phi_{\rm c}$ is visualized in Fig. \ref{fig:dist2d_1}(b).
	Here, this region is called the high-potential region.
	Incident ions cannot penetrate the high-potential region, and they are reflected to the $-x$-direction.
	Note that the high-density region does not exist in the result of the $\theta=\pi/2$ case.
	In the case of $\theta=0$, the size of the high-potential region in the $y$-direction, $L_{\rm c}$, is estimated as $L_{\rm c}=0.245$ m.
	The ion number flux flowing in the $x$-direction and approaching the high-potential region is given as
	\begin{equation}
		\Gamma_{\rm i,c}=n_0u_{\rm sat}L_{\rm c},
	\end{equation}
	where $\Gamma_{\rm i,c}$ is the ion number flux entering the high-potential region.
	Note that the unit length is considered in the $z$-direction.
	If one assumes that ions approaching the high-potential region are reflected to the $-x$-direction horizontally, the velocity of these ions changes from $+u_{\rm sat}$ to $-u_{\rm sat}$.
	Thus, the drag force generated by the reflected ions can be calculated from the momentum changes of these ions as follows:
	\begin{equation}
		F_{\rm c}=m_{\rm i}\Gamma_{\rm i,c}\cdot 2u_{\rm sat},
	\end{equation}
	where $F_{\rm c}$ is the estimated drag force generated by the reflected ions.
	$F_{\rm c}$ is calculated as 100.3 nN when $L_{\rm c}=0.245$ m.
	This value agrees well with the plasma drag force in the case of $\theta=0$.
	Therefore, the cause of the strong dependence of the plasma drag force on the magnetic moment angle is attributed to the strong drag force by the reflected ions as the case of $\theta=0$.
	
\begin{figure}[t]
		\begin{center}
   		\includegraphics[width=65mm]{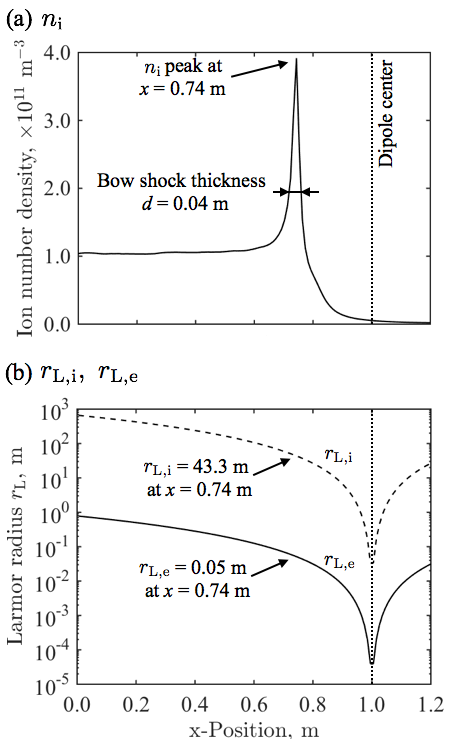}
		\end{center}
		\vspace{-2mm}
   	\caption{
		Distributions of the ion number density and Larmor radii of ions and electrons at $y=1.0$ m.}
   	\label{fig:dist1d_x}
   \end{figure}

	The electric field also plays an important role in the drag force generation in the case of $\theta=\pi/2$, even if there is no ion reflection.
	This fact can be derived by comparing the bow shock thickness with the Larmor radii of ions and electrons.
	Figure \ref{fig:dist1d_x}(a) shows the one-dimensional distribution of the ion number density $n_{\rm i}$ along the line of $y=1.0$ m in Fig. \ref{fig:dist2d_2}.
	$n_{\rm i}$ of the incoming flow is given by $n_0$, and it is discontinuously increased in the bow shock region.
	The plasma cavity exists behind the shock wave, and almost all the ions are evacuated around the MTQ dipole center.
	The peak of $n_{\rm i}$ is observed at $x=0.74$ m.
	If the boundary of the bow shock is defined by the contour line of $n_{\rm i}=2n_{0}$, the bow shock thickness $d$ is estimated as $d=0.04$ m.
	The distributions of the ion Larmor radius $r_{\rm L,i}$ and electron Larmor radius $r_{\rm L,e}$ are plotted in Fig. \ref{fig:dist1d_x}(b).
	Note that the magnetic field generated by the MTQ dipole is considered in this analysis.
	$r_{\rm L,i}$ and $r_{\rm L,e}$ at the location of the $n_{\rm i}$ peak are 43.3 m and 0.05 m, respectively.
	$r_{\rm L,e}$ is close to the bow shock thickness $d$, and hence electrons are magnetized at the location of the shock wave.
	On the other hand, $r_{\rm L,i}$ is greater than $d$ by two orders of magnitude.
	This means that ions are unmagnetized and mainly accelerated by electrostatic forces.

\subsection{Parametric Study on the Atmospheric Plasma Density and MTQ Magnetic Moment}
	\label{sec:para}
\begin{figure}[b]
   	\begin{center}
	   	\includegraphics[width=70mm]{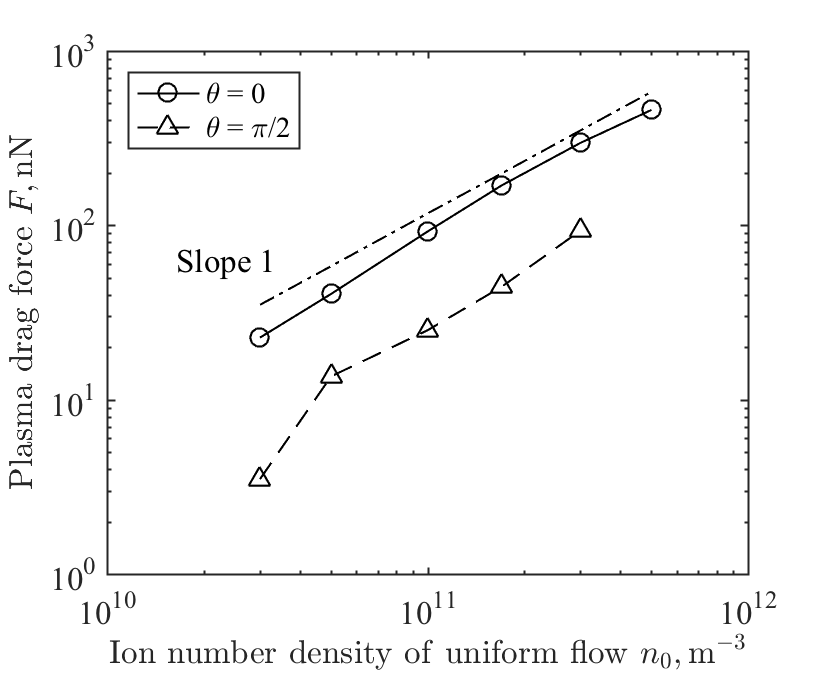}
   	\end{center}
		\vspace{-2mm}
		\caption{Dependence of the simulated plasma drag force on the plasma density of uniform flow.}
   	\label{fig:n0}
   \end{figure}

\begin{figure}[t]
   	\begin{center}
	   	\includegraphics[width=70mm]{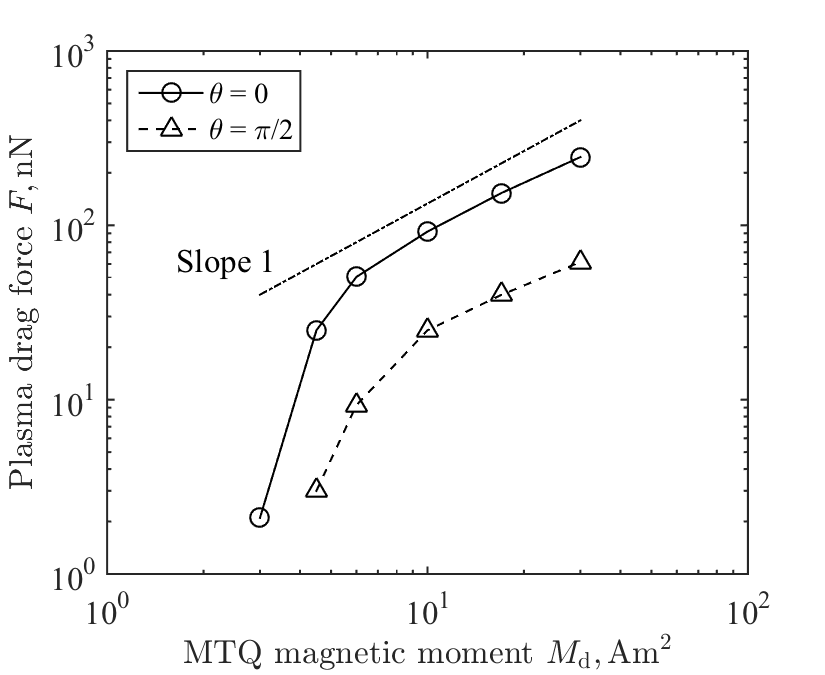}
   	\end{center}
		\vspace{-2mm}
		\caption{Dependence of the simulated plasma drag force on the magnetic moment of the MTQ.}
   	\label{fig:md}
   \end{figure}

	A parametric study was conducted to investigate the dependence of the plasma drag force on the atmospheric plasma density and magnetic moment of the MTQ.
	The dependence of the simulated plasma drag force on the atmospheric plasma density $n_{\rm 0}$ is shown in Fig. \ref{fig:n0}.
	In this analysis, the atmospheric plasma density in the orbit is varied between 0.3--3.0 $\times 10^{11}$ m$^{-3}$.
	This plasma density range corresponds to the altitude ranges of 550--1200 km during daytime and 250--650 km during nighttime.
	Additionally, the dependence of the plasma drag force on the MTQ magnetic moment $M_{\rm d}$ is shown in Fig. \ref{fig:md}.
	The magnetic moment is varied in the range of 3--30 A m$^{2}$.
	First, the relation of $F_{||}>F_{\perp}$ is always confirmed in this parametric study.
	In the researches on magnetic sails, the relation of $F_{||}<F_{\perp}$ is also observed in the cases of large $M_{\rm d}$, for which the ratios of the Larmor radius to the magnetosphere size for ions and electrons become $r_{\rm L,i}/L>1$ and $r_{\rm L,e}/L<1$, respectively [\citen{Kajimura:2012aa}].
	However, $M_{\rm d}$ of 30 A m$^2$ is already large for the MTQs in nanosatellites, and it would be difficult to use a sufficiently large $M_{\rm d}$ to make the conditions of $r_{\rm L,e}<L$ and $r_{\rm L,i}>L$.
	Therefore, the relation of $F_{||}>F_{\perp}$ is supposed to be valid within the range of $M_{\rm d}$ used in the magnetic plasma deorbit method.
	Concerning the $F_{||}$, the relation of $F_{||}\propto n_0^{1}$ is confirmed, and $F_{||}\propto M_{\rm d}^{1}$ is also observed when $M_{\rm d}>10$ A m$^{2}$.
	According to the analytic model in Eqs. (\ref{eq:model1}) and (\ref{eq:model2}), the plasma drag force is supposed to change according to the relations of $F\propto n_{\rm 0}^{2/3}$ and $F\propto M_{\rm d}^{2/3}$.
	Thus, the analytic model is not valid in predicting the effect of variations in $n_{\rm 0}$ and $M_{\rm d}$ on the plasma drag force in the magnetic plasma deorbit.
	Note that Ashida et al. also showed that $F$ is proportional to $M_{\rm d}$ in the cases of small-scale magnetic sails where finite-Larmor radius effects become significant [\citen{Ashida:2014aa}].


\subsection{Deorbit Analysis}
\begin{figure}[t]
   	\begin{center}
	   	\includegraphics[width=70mm]{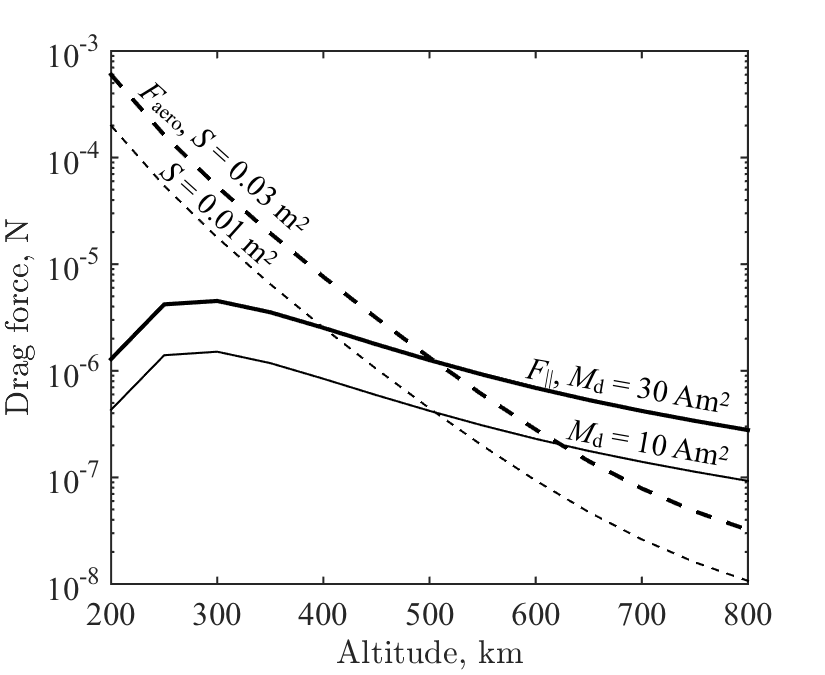}
   	\end{center}
		\vspace{-2mm}
		\caption{Dependences of the aerodynamic and plasma drag forces on the altitude.}
   	\label{fig:Falt}
   \end{figure}

	The deorbit duration achieved by the magnetic plasma deorbit method is estimated by using the plasma drag forces obtained by the present simulation.
	First, the effect of the aerodynamic force is considered.
	The aerodynamic drag force $F_{\rm aero}$ is generally formulated as follows:
	\begin{equation}
		F_{\rm aero}=\frac{1}{2}C_{\rm d,aero}\rho_0 u_{\rm sat}^2S.
	\end{equation}
	The atmospheric gas density $\rho_0$ at each altitude can be calculated by using the NRLMSISE-00 atmosphere model [\citen{PiconeJGR2002}].
	Numerous models have been proposed to model the drag coefficient $C_{\rm d,aero}$ [\citen{MOE:1993aa,Mostaza-Prieto:2014aa,Vallado:2014aa}].
	It is known that a satellite of a relatively simple shape like a sphere has a drag coefficient ranging 2.0--2.4 in the LEO [\citen{Moe:1998aa}].
	Here, a constant drag coefficient of $C_{\rm d,aero}=2.2$ is assumed for simplicity.
	Concerning the plasma drag force, the case of the magnetic moment parallel to the satellite velocity is considered, i.e., $F_{||}$ is assumed as the plasma drag force.
	Figure \ref{fig:Falt} shows the dependences of the plasma and aerodynamic drag forces on the altitude.
	$F_{\rm aero}$ with face areas $S$ of $0.03$ m$^2$ and $0.01$ m$^2$ are plotted, whereas $F_{||}$ with MTQ magnetic moments $M_{\rm d}$ of $30$ A m$^2$ and $10$ A m$^2$ are shown.
	Both of $F_{\rm aero}$ and $F_{||}$ are weakened as the altitude becomes higher, because both the atmospheric gas and plasma densities decrease exponentially with the altitude.
	At a relatively low altitude of $h<$ 500 km, the aerodynamic force is predominant.
	However, $F_{\rm aero}$ decreases more rapidly with the altitude than $F_{||}$, and $F_{||}$ becomes comparable with $F_{\rm aero}$ within 500 km $< h <$ 600 km.
	At an altitude exceeding 600 km, the plasma drag force is more effective than the aerodynamic force.
	Therefore, the magnetic plasma deorbit method is expected to be beneficial for the satellite deorbit from high altitudes of $h>$ 600 km.
	
\begin{figure}[t]
   	\begin{center}
	   	\includegraphics[width=70mm]{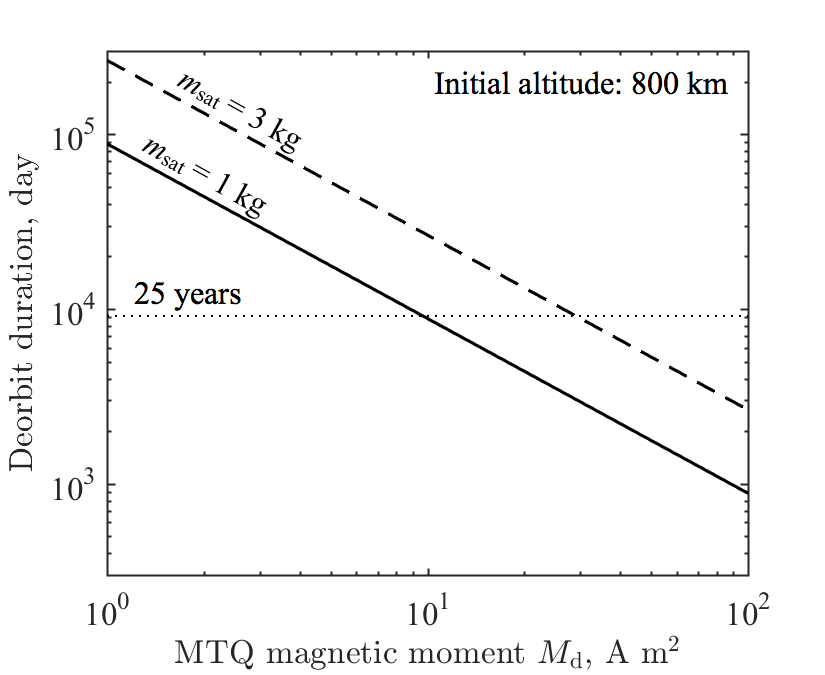}
   	\end{center}
		\vspace{-2mm}
		\caption{Deorbit duration attained by the magnetic plasma deorbit for satellites at an altitude of 800 km.}
   	\label{fig:Duration}
   \end{figure}

	An analytic model to predict the operation period of the magnetic plasma deorbit method is proposed in Ref. [\citen{Inamori2015192}].
	In this model, the plasma density dependency on the altitude is approximated as follows:
	\begin{equation}
		\rho = \rho_{\rm ref}\exp\left(-\frac{\sqrt{a}-\sqrt{a_{\rm ref}}}{H}\right),
		\label{eq:duration0}
	\end{equation}
	where $H$ is a constant fitting parameter, which is assumed to be $H=25$ m$^{\frac{1}{2}}$ [\citen{Inamori2015192}].
	The satellite semi-major axis is expressed by a perturbation equation for a circular orbit.
	The deorbit duration is obtained by analytically integrating the perturbation equation as follows:
	\begin{equation}
		t_{\rm deorbit} = \frac{2m_{\rm sat}H}{C_{\rm d,p}\rho_{\rm ref}\pi L^2\sqrt{\gamma}}G,
		\label{eq:duration1}
	\end{equation}
	where 
	\begin{equation}
		G = \exp\left(\frac{\sqrt{a_{\rm ini}}-\sqrt{a_{\rm ref}}}{H}\right)
		   -\exp\left(\frac{\sqrt{a_{\rm E}}-\sqrt{a_{\rm ref}}}{H}\right).
		\label{eq:duration2}
	\end{equation}	
	$\gamma$ and $a_{\rm E}$ are the standard gravitational parameter and Earth's radius, respectively.
	The detailed derivation process of the deorbit duration is explained in Ref. [\citen{Inamori2015192}].
	Only the plasma drag force is taken into consideration in this model, and Eq. (\ref{eq:duration1}) gives an instantaneous estimation of the deorbit duration.
	By using Eq. (\ref{eq:model1}), Eq. (\ref{eq:duration1}) can be rewritten with the plasma drag force $F$.
	Here, $F_{||}$ is assumed as the plasma drag force, and $F_{||}\propto M_{\rm d}$ is further assumed according to the discussion in Sec. IV.C.
	Owing to these assumptions, the plasma drag force is expressed as $F=F_{\rm ||,ref}M_{\rm d}/M_{\rm d,ref}$, and Eq. (\ref{eq:duration1}) is rewritten as follows:
	\begin{equation}
		t_{\rm deorbit} = \frac{M_{\rm d,ref}Hu_{\rm sat}^2}{F_{\rm ||,ref}\sqrt{\gamma}}\frac{m_{\rm sat}}{M_{\rm d}}G.
		\label{eq:duration3}
	\end{equation}
	Note that $u_0$ in Eq. (\ref{eq:model1}) is replaced by $u_{\rm sat}$ in the case of the magnetic plasma deorbit.
	The reference values are determined based on the condition of $h=800$ km, as follows: $a_{\rm ref}=7180$ km, $M_{\rm d,ref}=10$ A m$^2$, and $F_{\rm ||,ref}=105$ nN.
	A constant satellite velocity is assumed as $u_{\rm sat}=8,000$ m s$^{-1}$.
	The deorbit duration is estimated for 1 kg and 3 kg satellites with the initial altitude of 800 km. 
	The results are plotted with various MTQ magnetic moments in Fig. \ref{fig:Duration}.
	For instance, $M_{\rm d}$ of 30 A m$^2$ is required for the deorbit of a 3 kg satellite within 25 y.
	The effectiveness of the magnetic plasma deorbit method is quantitatively evaluated by this analysis.
	In order to predict the deorbit duration more accurately, one needs to consider several effects, such as the satellite attitude, geomagnetic field, finite satellite body and coil, hour of day, and so on.
	In addition, the optimization of the deorbit scenario considering these effects would also be one of the themes of subsequent papers.
	
\section{Conclusion}
	A plasma flow simulation was conducted to predict plasma drag forces generated by the magnetic plasma deorbit method.
   Static electric and magnetic fields are assumed, and the plasma flow is described by using the fully kinetic model.
   The simulation is continued until a steady state is reached, and the steady plasma drag force is calculated.
   The simulated plasma drag forces are compared with those predicted by the analytic model of magnetic sails, and the validity of the model is examined. 
   The findings are summarized as follows:
   \begin{enumerate}
      \item The structures of the bow shock and magnetosphere are observed in the vicinity of the MTQ, and a steady plasma drag force is generated.
      The basic concept of the drag force generation is confirmed.
      \item The predicted $F$ is 105.2 nN in the case of $n_0=1.0\times 10^{11}$ m$^{-3}$, $u_0=8.0$ km s$^{-1}$, $M_{\rm d}=10$ A m$^2$, and $\theta=0$.
      $C_{\rm d,p}$ is estimated as 0.2. 
      This small $C_{\rm d,p}$ indicates that the analytic model of magnetic sails is not applicable to the prediction of plasma drag forces in the magnetic plasma deorbit.
      \item The plasma drag force shows a strong dependence on the MTQ magnetic moment angle.
      The drag force in the case of the magnetic moment parallel to the satellite velocity (i.e., $F_{||}$) is 3.5 times larger than that in the case of the magnetic moment perpendicular to the satellite velocity (i.e., $F_{\perp}$).
      This strong dependence is attributed to a strong drag force in the case of $\theta=0$, which is mainly generated by reflected ions.
      The relation of $F_{||}>F_{\perp}$ is valid within the conditions of the magnetic plasma deorbit method.
      \item The dependence of the plasma drag force in the case of the magnetic moment parallel to the satellite velocity ($F_{||}$) on the atmospheric plasma density and the MTQ magnetic moment was linear.
   \end{enumerate}

\clearpage
\section*{Appendix A: Validity of the AEM Model}
	\label{sec:app1}
\begin{figure}[t]
   	\begin{center}
	   	\includegraphics[width=70mm]{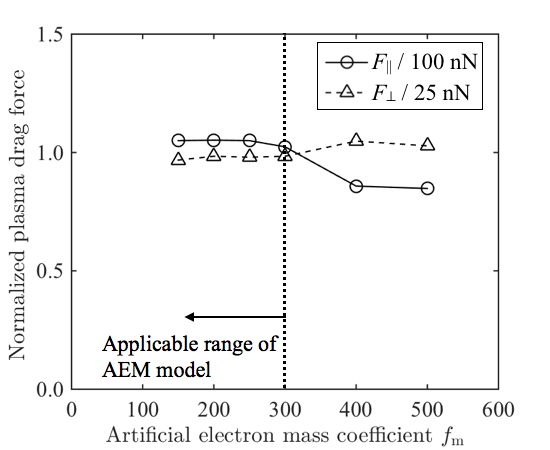}
   	\end{center}
		\vspace{-2mm}
		\caption{Dependence of the simulated plasma drag force on the AEM coefficient.}
   	\label{fig:aem}
   \end{figure}

	As stated in Sec. \ref{sec:aem}, the AEM model is used in the simulation to mitigate the issue of numerical stiffness.
	The validity of the AEM model is examined by confirming the convergence of the plasma drag force when the AEM coefficient $f_{\rm m}$ is changed from 150 to 500.
	$F_{||}$ ($\theta = 0$ case) and $F_{\perp}$ ($\theta = \pi/2$ case) calculated with various $f_{\rm m}$ are shown in Fig. \ref{fig:aem}.
	In this figure, $F_{||}$ and $F_{\perp}$ are normalized by 100 nN and 25 nN, respectively.
	Note that $f_{\rm m}=1$ means that the real electron mass is used and the electron motion approaches the real situation. 
	In the case of $f_{\rm m}=150$, $F_{||}$ and $F_{\perp}$ are 105.1 nN and 24.6 nN, respectively.
	As $f_{\rm m}$ is increased, the calculation time is reduced, but the difference between the AEM and the real electron mass becomes large.
	In the case of $f_{\rm m}=300$, $F_{||}$ and $F_{\perp}$ are 102.4 nN and 24.1 nN, respectively.
	The differences in the plasma drag forces between the cases of $f_{\rm m}=300$ and $f_{\rm m}=150$ are $-2.7\%$ and $-0.4\%$ for $F_{||}$ and $F_{\perp}$, respectively.
	Therefore, the variations in $F_{||}$ and $F_{\perp}$ are considered to be small when $1\leq f_{\rm m} \leq 300$.
	It is important to find a point of compromise that satisfies both reasonable calculation time and accuracy.
	In the simulation of the magnetic plasma deorbit, $f_{\rm m}=200$ is selected based on several test calculations.
	
\section*{Appendix B: Model Validation by a Magnetic Sail Simulation}
	To check the validity of the present model, the thrust of a magnetic sail is simulated, and the result is compared with previous numerical simulation results.
	A magnetic sail is one of the propulsion systems for interplanetary transportation originally proposed by Zubrin and Andrews [\citen{ZUBRIN:1991aa}].
	This system obtains its thrust by the interaction between solar wind plasma and the magnetic field generated by an onboard magnetic coil.
	Although the plasma properties of the solar wind are different from those of the LEO plasma, a numerical simulation of a magnetic sail would yield a quantitative evaluation regarding the adequacy of the present model.
	
	As written in Sec. II.A., the plasma flow surrounding a dipole magnetic field can be characterized by the ratio of the Larmor radius $r_{\rm L}$ to the magnetosphere size $L$.
	In the case of the magnetic plasma deorbit, the criteria for ions and electrons are $r_{\rm L,i}/L>1$ and $r_{\rm L,e}/L>1$, respectively.
	The thrust of magnetic sails under the conditions of $r_{\rm L,i}/L>1$ and $r_{\rm L,e}/L>1$ was evaluated by three-dimensional full particle simulations in Ref. [\citen{Ashida:2014aa}].
	The properties of the solar wind assumed in this simulation are as follows: plasma density is 5$\times$10$^6$ m$^{-3}$, solar wind velocity is 5$\times$10$^5$ m s$^{-1}$, plasma temperature is 10 eV, and all ions are assumed to be H$^+$.
	The interplanetary magnetic field is ignored, and the dipole magnetic moment is assumed.
	
   \begin{figure}[t]
   	\begin{center}
	   	\includegraphics[width=80mm]{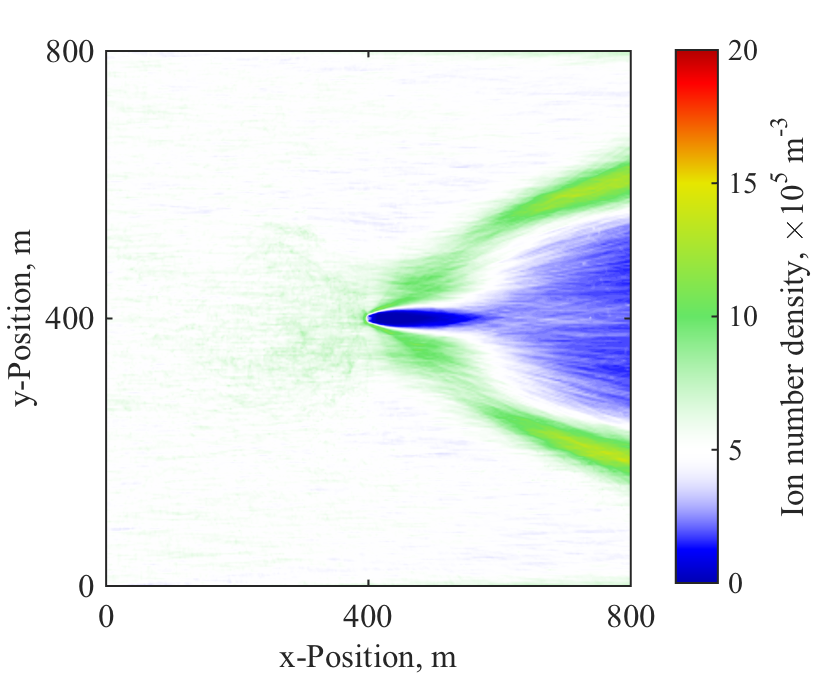}
   	\end{center}
		\vspace{-2mm}
		\caption{2D distribution of the ion number density in the case of $M_{\rm d}=10^5$ A m$^{2}$.}
   	\label{fig:ni_solar}
   \end{figure}
   
	A magnetic sail with a dipole of $M_{\rm d}=10^5$ A m$^{2}$ is simulated by using the present model in the calculation domain of 800 m$\times$800 m with a grid of 300$\times$300.
	The dipole is located at the center of the domain and with a magnetic moment angle of $\theta=\pi/2$.
	The 2D distribution of the ion number density is shown in Fig. \ref{fig:ni_solar}.
	The solar wind is deflected by the magnetic field, and a plasma cavity is formed behind the dipole.
	The simulated thrust $F_{\rm sim}$ at the steady state is given as $F_{\rm sim}=58.9$ nN.
	An empirical equation to predict the thrust of a small-scale magnetic sail is presented based on the results of the three-dimensional simulations as follows [\citen{Ashida:2014aa}]:
	\begin{equation}
		\log_{10}F_{\rm emp}=1.03\log_{10}M_{\rm d}-12.5.
		\label{eq:magsail}
	\end{equation}
	In the case of $M_{\rm d}=10^5$ A m$^{2}$, the thrust predicted by this empirical equation is $F_{\rm emp}=44.7$ nN.
	The difference between $F_{\rm sim}$ and $F_{\rm emp}$ is 14.2 nN, which is 24\% of $F_{\rm sim}$.
	Thus, the simulation using the present model may include this level of uncertainty.
	Note that this comparison is made between the present 2D simulation and the previous three-dimensional simulations.
	A more suitable validation method is to compare simulation results with experimental data. 
	This is one of the themes of future works.

\bibliography{reference}

\begin{thebibliography}{10}
\newcommand{\enquote}[1]{``#1''}

\bibitem{SwartwoutJoSS2013}
Swartwout, M., \enquote{The First One Hundred CubeSats: A Statistical Look,}
  {\em Journal of Small Satellites\/}, Vol.~2, No.~2, 2013, pp.~213--233.

\bibitem{Funase:2007aa}
Funase, R., Takei, E., Nakamura, Y., Nagai, M., Enokuchi, A., Yuliang, C.,
  Nakada, K., Nojiri, Y., Sasaki, F., Funane, T., Eishima, T., and Nakasuka,
  S., \enquote{Technology demonstration on University of Tokyo's pico-satellite
  ``XI-V''and its effective operation result using ground station network,}
  {\em Acta Astronautica\/}, Vol.~61, No.~7, 2007, pp.~707--711.

\bibitem{Inamori2015192}
Inamori, T., Kawashima, R., Saisutjarit, P., Sako, N., and Ohsaki, H.,
  \enquote{Magnetic plasma deorbit system for nano- and micro-satellites using
  magnetic torquer interference with space plasma in low Earth orbit,} {\em
  Acta Astronautica\/}, Vol.~112, 2015, pp.~192--199.

\bibitem{ZUBRIN:1991aa}
Zubrin, R.~M. and Andrews, D.~G., \enquote{Magnetic sails and interplanetary
  travel,} {\em Journal of Spacecraft and Rockets\/}, Vol.~28, No.~2, 1991,
  pp.~197--203.

\bibitem{Funaki2007}
Funaki, I., Kojima, H., Yamakawa, H., Nakayama, Y., and Shimizu, Y.,
  \enquote{Laboratory Experiment of Plasma Flow Around Magnetic Sail,} {\em
  Astrophysics and Space Science\/}, Vol.~307, No.~1, 2007, pp.~63--68.

\bibitem{Nishida:2012aa}
Nishida, H. and Funaki, I., \enquote{Analysis of Thrust Characteristics of a
  Magnetic Sail in a Magnetized Solar Wind,} {\em Journal of Propulsion and
  Power\/}, Vol.~28, No.~3, 2012, pp.~636--641.

\bibitem{Kajimura:2012aa}
Kajimura, Y., Funaki, I., Matsumoto, M., Shinohara, I., Usui, H., and Yamakawa,
  H., \enquote{Thrust and Attitude Evaluation of Magnetic Sail by
  Three-Dimensional Hybrid Particle-in-Cell Code,} {\em Journal of Propulsion
  and Power\/}, Vol.~28, No.~3, 2012, pp.~652--663.

\bibitem{Ashida:2013aa}
Ashida, Y., Funaki, I., Yamakawa, H., Usui, H., Kajimura, Y., and Kojima, H.,
  \enquote{Two-Dimensional Particle-In-Cell Simulation of Magnetic Sails,} {\em
  Journal of Propulsion and Power\/}, Vol.~30, No.~1, 2013, pp.~233--245.

\bibitem{Harned:1982aa}
Harned, D.~S., \enquote{Quasineutral hybrid simulation of macroscopic plasma
  phenomena,} {\em Journal of Computational Physics\/}, Vol.~47, No.~3, 1982,
  pp.~452--462.

\bibitem{Kajimura:2010aa}
Kajimura, Y., Usui, H., Funaki, I., Ueno, K., Nunami, M., Shinohara, I.,
  Nakamura, M., and Yamakawa, H., \enquote{Hybrid Particle-in-Cell Simulations
  of Magnetic Sail in Laboratory Experiment,} {\em Journal of Propulsion and
  Power\/}, Vol.~26, No.~1, 2017/12/20 2010, pp.~159--166.

\bibitem{AshidaThesis}
Ashida, Y., \enquote{Study on Propulsive Characteristics of Magnetic Sail and
  Magneto Plasma Sail by Plasma Particle Simulations,} Ph.D. Thesis, Kyoto
  University, 2014.

\bibitem{chen1984introduction}
Chen, F.~F., {\em Introduction to plasma physics and controlled fusion\/},
  Plenum Press, New York, 2nd ed., 1984.

\bibitem{ChoPoP2013}
Cho, S., Komurasaki, K., and Arakawa, Y., \enquote{Kinetic particle simulation
  of discharge and wall erosion of a {H}all thruster,} {\em Physics of
  Plasmas\/}, Vol.~20, No.~6, 2013, pp.~063501.

\bibitem{Szabo:2013aa}
Szabo, J., Warner, N., Martinez-Sanchez, M., and Batishchev, O., \enquote{Full
  Particle-In-Cell Simulation Methodology for Axisymmetric {H}all Effect
  Thrusters,} {\em Journal of Propulsion and Power\/}, Vol.~30, No.~1, 2013,
  pp.~197--208.

\bibitem{SchunkBook2004}
Schunk, R. and Nagy, A., {\em Ionospheres: physics, plasma physics, and
  chemistry\/}, Cambridge University Press, 2009.

\bibitem{Bilitza}
Bilitza, D., \enquote{International Reference Ionosphere 1990,} {\em NASA
  Technical Report\/}, 1990, pp.~0--84.

\bibitem{HastingsJGR1995}
Hastings, D.~E., \enquote{A review of plasma interactions with spacecraft in
  low Earth orbit,} {\em Journal of Geophysical Research: Space Physics\/},
  Vol.~100, No.~A8, 1995, pp.~14457--14483.

\bibitem{BirdsallBook2004}
Birdsall, C.~K. and Langdon, A.~B., {\em Plasma physics via computer
  simulation\/}, CRC Press, 2004.

\bibitem{MASON1981233}
Mason, R.~J., \enquote{Implicit moment particle simulation of plasmas,} {\em
  Journal of Computational Physics\/}, Vol.~41, No.~2, 1981, pp.~233--244.

\bibitem{Nishida:2006aa}
Nishida, H., Ogawa, H., Funaki, I., Fujita, K., Yamakawa, H., and Nakayama, Y.,
  \enquote{Two-Dimensional Magnetohydrodynamic Simulation of a Magnetic Sail,}
  {\em Journal of Spacecraft and Rockets\/}, Vol.~43, No.~3, 2006,
  pp.~667--672.

\bibitem{Ashida:2014aa}
Ashida, Y., Yamakawa, H., Funaki, I., Usui, H., Kajimura, Y., and Kojima, H.,
  \enquote{Thrust Evaluation of Small-Scale Magnetic Sail Spacecraft by
  Three-Dimensional Particle-in-Cell Simulation,} {\em Journal of Propulsion
  and Power\/}, Vol.~30, No.~1, 2014, pp.~186--196.

\bibitem{PiconeJGR2002}
Picone, J.~M., Hedin, A.~E., Drob, D.~P., and Aikin, A.~C.,
  \enquote{NRLMSISE-00 empirical model of the atmosphere: Statistical
  comparisons and scientific issues,} {\em Journal of Geophysical Research:
  Space Physics\/}, Vol.~107, No. A12, 2002, pp.~SIA 15--1--SIA 15--16, 1468.

\bibitem{MOE:1993aa}
Moe, M.~M., Wallace, S.~D., and Moe, K., \enquote{Refinements in determining
  satellite drag coefficients - Method for resolving density discrepancies,}
  {\em Journal of Guidance, Control, and Dynamics\/}, Vol.~16, No.~3, 1993,
  pp.~441--445.

\bibitem{Mostaza-Prieto:2014aa}
Mostaza~Prieto, D., Graziano, B.~P., and Roberts, P. C.~E., \enquote{Spacecraft
  drag modelling,} {\em Progress in Aerospace Sciences\/}, Vol.~64, No.
  Supplement C, 2014, pp.~56--65.

\bibitem{Vallado:2014aa}
Vallado, D.~A. and Finkleman, D., \enquote{A critical assessment of satellite
  drag and atmospheric density modeling,} {\em Acta Astronautica\/}, Vol.~95,
  No. Supplement C, 2014, pp.~141--165.

\bibitem{Moe:1998aa}
Moe, K., Moe, M.~M., and Wallace, S.~D., \enquote{Improved Satellite Drag
  Coefficient Calculations from Orbital Measurements of Energy Accommodation,}
  {\em Journal of Spacecraft and Rockets\/}, Vol.~35, No.~3, 1998,
  pp.~266--272.

\end{thebibliography}
\bibliographystyle{aiaa}

\end{document}